\documentclass[twocolumn,prb,showpacs,superscriptaddress]{revtex4-1}

\usepackage{graphicx}
\begin{document}

\title{Interplay of disorder and geometrical frustration in doped Gadolinium Gallium Garnet}

\author{N. Woo}
\affiliation{The James Franck Institute and Department of Physics, The University of Chicago, Chicago, IL 60637, USA}
\author{D.M. Silevitch}
\affiliation{The James Franck Institute and Department of Physics, The University of Chicago, Chicago, IL 60637, USA}
\author{C. Ferri}
\affiliation{Department of Physics, University of California, Merced, CA 95343, USA}
\author{S. Ghosh}
\affiliation{Department of Physics, University of California, Merced, CA 95343, USA}
\author{T.F. Rosenbaum}
\email[email: ]{tfr@caltech.edu}
\affiliation{The James Franck Institute and Department of Physics, The University of Chicago, Chicago, IL 60637, USA}
\affiliation{Division of Physics, Math and Astronomy, California Institute of Technology, Pasadena, CA 91125, USA}
\thanks{}

\date{\today}
\begin{abstract}
The geometrically-frustrated, triangular antiferromagnet GGG exhibits a rich mix of short-range order and isolated quantum states. We investigate the effects of up to 1\% Neodymium substitution for Gallium on the ac magnetic response at temperatures below 1 K in both the linear and nonlinear regimes. Substitutional disorder actually drives the system towards a more perfectly frustrated state, apparently compensating for the effect of imperfect Gadolinium/Gallium stoichiometry, while at the same time more closely demarcating the boundaries of isolated, coherent clusters composed of hundreds of spins. Optical measurements of the local Nd environment substantiate the picture of an increased frustration index with doping.
\end{abstract}

\pacs{75.50.Lk, 75.40.Gb, 75.45.+j, 75.30.Hx}
\maketitle

\begin{text}

Geometrically frustrated magnets suppress the ability of spins to freeze to temperatures well below their natural ordering scales and, in so doing, open to experimental scrutiny low temperature fluctuations, extended quantum states, and thermodynamically isolated degrees of freedom. The topology of the interspin interactions makes it impossible to simultaneously minimize all of the pairwise interaction energies \cite{Ramirez94} and, in principle, a fully frustrated system cannot form a long-range ordered state, even at absolute zero. Short-range effects can dominate the macroscopic behavior, with self-organization of spins into such quantities as quantum protectorates: coherent states that are topologically isolated from many sources of environmental decoherence.\cite{Laughlin00,Lee02} The introduction of structural or compositional disorder to such a frustrated system has the potential to tune the frustration effects and their macroscopic expressions. It is not clear \textit{a priori}, however, whether additional disorder enhances self-organization by further isolating the spin clusters, or actually relieves frustration by breaking the symmetry of the interspin couplings and hence degrading the isolation between any extended quantum states and the environment.\cite{Villain80,Shender82} Here, we study the effects of progressively increasing the degree of disorder in a geometrically frustrated antiferromagnet by introducing controlled amounts of a magnetic dopant.

The parent compound, Gadolinium Gallium Garnet (Gd$_{3}$Ga$_{5}$O$_{12}$ or GGG), has a cubic lattice in which the magnetic Gd$^{3+}$ ions are located on two interpenetrating sub-lattices composed of corner sharing triangles with a Heisenberg spin symmetry and a single ion anisotropy of less than 0.04 K.\cite{Overmeyer83} Nearest neighbors are connected by corner sharing triangles on each sub-lattice, coupled by a 1.5 K antiferromagnetic exchange interaction and a 0.7 K dipolar interaction, and the two sub-lattices interact via next-nearest-neighbor interactions. The resulting Weiss temperature is of order 2.3 K.\cite{Kinney79} The combination of the antiferromagnetic interaction and the topology of couplings provides a macroscopic number of equally-probable spin configurations at absolute zero.\cite{Lacroix11}

Several experimental studies of GGG have probed the subtle nature of the ground state, with differing results. Early bulk thermodynamic measurements \cite{Schiffer95} found evidence of spin freezing and an associated spin glass transition at $T \sim 130$ mK. By contrast, muon spin relaxation \cite{Dunsiger00,Marshall02} and Mossbauer \cite{Bonville04} measurements found that the spins remained unfrozen to temperatures as low as 25 mK. Powder neutron measurements observed short-range correlations at elevated temperatures followed by a 140 mK transition to a state with longer range correlations and spin-liquid type behavior.\cite{Petrenko98} Subsequently, a combination of linear and nonlinear ac susceptibility measurements found a state consisting of a set of ``quantum protectorates'', extended coherent states that are topologically decoupled from the background, coexisting with at least short-range antiferromagnetic order onsetting at $T  \sim 100$ mK.\cite{Ghosh08} More recent high-resolution heat capacity measurements\cite{Quilliam13} did not observe any evidence of a spin glass transition in the 120--200 mK range. The interesting spin dynamics all occur at temperatures below 200 mK, over an order of magnitude lower than the Weiss temperature, confirming the picture of a highly frustrated spin configuration.

One effect contributing to this range of results is the difficulty in growing purely stoichiometric crystals of GGG, as excess magnetic Gd$^{3+}$ ions tend to randomly substitute for nonmagnetic Ga$^{3+}$ ions.\cite{Daudin82,Schiffer94}  Gadolinium ions occupying gallium sites have a nearest-magnetic-neighbor distance that is 9\% smaller than the normal Gd-Gd distance in GGG, resulting in both lattice strain and a local increase in the magnetic interaction strength.\cite{Quilliam13} Against this backdrop, we examine a series of GGG crystals with up to 1\% of Nd${}^{3+}$ dopants substituting for Gd$^{3+}$ ions. The Nd ions introduce a controlled degree of disorder into the magnetic lattice that breaks the local symmetry of the magnetic interactions. By controlling the level of this quenched disorder, we can study the role it plays in the formation and evolution of the low-temperature magnetic phases.

We performed ac susceptibility measurements on a series of ($5\times 5 \times 10$) mm$^{3}$ single crystals of GGG:Nd$_x$, with  $x = 0$, 0.1, and 1.0\% (Princeton Scientific). X-ray lattice constant measurements showed that all three crystals exhibited a similar degree of off-stoichiometry of approximately 3\%. We employed a gradiometric susceptometer attached to the cold finger of a helium dilution refrigerator, using sapphire rods pressed against the faces of the crystals to provide a thermal link to the cryostat. The complex ac susceptibility was measured in the linear-response limit using 40 mOe of probe field and mapped as functions of frequency (1 Hz to 10 kHz) and temperature (30 to 300 mK) for all three dopings.

The nonlinear response of the system provides the most acute insight into the behavior of the local, isolated spin clusters. To that end, we measured the susceptibility as a function of drive amplitude (mOe to Oe) at fixed frequency. In addition to the direct effect on the GGG crystals, large ac magnetic fields, particularly at kilohertz frequencies, also induce eddy currents in metallic mounts, resulting in significant Ohmic heating. To alleviate this issue, we mounted the susceptometer on a carbon-fiber/sapphire framework \cite{Schmidt13} designed to minimize eddy currents while maintaining a high thermal conductivity for efficient thermal linkage to the cryostat. The ac probe field was applied using a duty cycle of 1--2\% to further reduce the heat load.

\begin{figure}
\includegraphics[clip,trim=100 420 100 55,width=250pt]{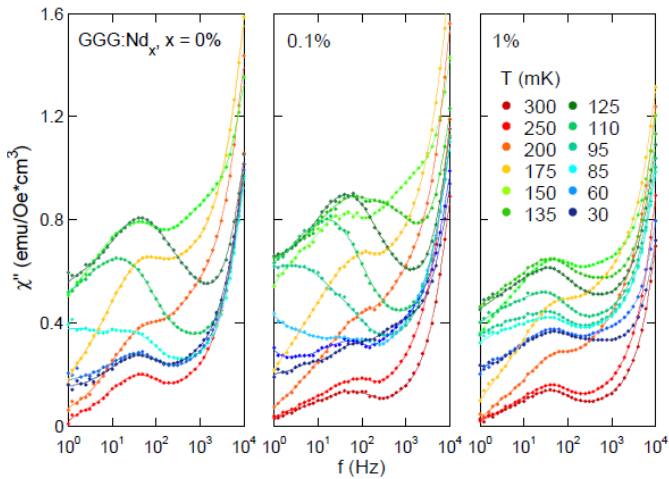}
\caption{Imaginary part of the ac magnetic susceptibility $\chi''$ as a function of frequency $f$ at a series of temperatures $T$ for GGG:Nd$_x$ with $x = 0$, 0.1\%, and 1\%. The susceptibility was measured using a 40 mOe probe field to ensure linear response. A low-frequency mode at 50 Hz and a high-frequency mode that peaks above the 10 kHz measurement window characterize the global response of all three materials. For $x = 0$ and 0.1\%, a flat low-frequency response develops at 85 to 95 mK, corresponding to fluctuations associated with ordering. At $x= 1.0$\%, this feature is not observed, indicating that any ordering is suppressed below 30 mK.}
\label {fig1}
\end{figure}

Finally, we exploited the coupling of the Nd ions to the Gd ions to probe optically the local magnetic environment.  Lifetime data was calculated from the homogenous linewidth of the $R_1 \to Z_5$ transition of the ${}^4F_{3/2} \to {}^4I_{9/2}$ ground to first excited state emission multiplets of the Nd$^{3+}$ ion. The samples were mounted in a liquid helium flow cryostat and excited with a Ti:Sapphire laser tuned to 808 nm. The excitation wavelength was chosen to be as close to the absorption band edge as possible for the ${}^4F_{3/2} \to {}^4I_{9/2}$ emission lines. The emission is collected in reflection and dispersed by an Acton 300i spectrometer onto a thermoelectrically cooled CCD with spectral resolution of 0.18 nm.

\begin {figure}
\includegraphics[clip,trim=140 440 120 60,width=250pt]{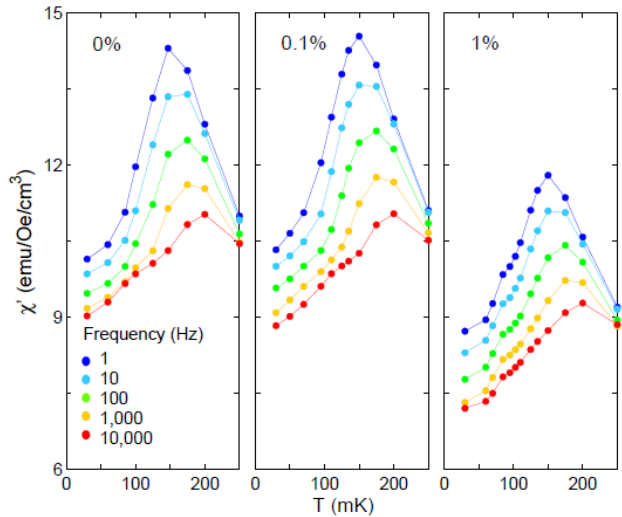}
\caption{Real part of the susceptibility $\chi'$ as a function of temperature $T$ at a series of frequencies $f$ for GGG:Nd$_x$ with $x = 0$, 0.1\%, and 1\%. The functional form and temperature scale for all three doping levels is similar; however, the static incoherent background is suppressed for $x = 1.0$\% as frustration is enhanced by doping.}
\label {fig2}
\end {figure}
The complex ac susceptibility at audio frequencies delineates the nature and timescales of the low-energy magnetic modes of the material. Peaks in the imaginary component of the susceptibility, $\chi''$, yield information on characteristic relaxation times. The presence of a flat response in the $f \to 0$ limit corresponds to $1/f$ noise in the magnetization, a signature of a spin state with scale invariance and hence a complex hierarchy of energies.\cite{Ocio86,Fisher86} We plot in Fig. \ref{fig1} imaginary susceptibility spectra for GGG:Nd$_x$ ($x=0$, 0.1, 1\%) for $30 \leq T \leq 300$ mK. For all three samples, we observe two peaks in the susceptibility, one at 50 Hz and one that appears to lie above the 10 kHz upper limit of the measurement. This corresponds to two characteristic relaxation timescales in the system, one at 20 msec and one shorter than 100 $\mu$sec.

The characteristic long-time relaxation at 50 Hz does not depend strongly on either $T$ or $x$. By contrast, the behavior of the system in the low-frequency limit changes markedly with $x$.  For pure GGG and GGG with 0.1\% Nd, we observe a clear $f \to 0$ plateau in the imaginary susceptibility at $T \sim 85$ to 95 mK. No such plateau appears in the 1.0\% doped material. The appearance of such a plateau is often correlated with the entry into a spin glass state.\cite{Ocio86,Wu91} If that were the case here, the plateau should remain robust for all $T$ below the glass transition temperature. Instead, the plateau behavior is only stable over a finite temperature band, below which the susceptibility approaches zero with finite slope. We associate the plateau with antiferromagnetic ordering, at least on the 100 $\mathring{A}$ scale.\cite{Petrenko98} The introduction of Nd suppresses the ordering to $T < 30$ mK, effectively restoring a higher degree of geometric frustration.

The relative frustration index as a function of doping also can be discerned by looking at the real component, $\chi'$, of the linear susceptibility, as shown in Fig. \ref{fig2} for all three sample concentrations. $\chi'$ consists of a frequency-dependent peaked component riding on a frequency independent background. The peak reflects the freezing out of magnetic modes at the pertinent time scale of the interrogation frequency, but conspicuously does not go to zero (note the suppressed zero of Fig. \ref{fig2}). We attribute this background to a Heisenberg spin bath made available by imperfections in the crystal and deviations from stoichiometry. Rather than increasing the background susceptibility, the Nd ions at the highest level of substitution effectively reduce the degree of disorder and push the system towards a more perfectly frustrated state.

We can ask whether this apparently counter-intuitive result also applies to the quantum protectorates previously reported for undoped GGG.\cite{Ghosh08} An experimental signature of these coherent spin clusters in GGG occurs in the nonlinear susceptibility. As a function of drive field amplitude, the magnetization traces out a Brillouin function, with approach to saturation above 1 Oe. This behavior is characteristic of clusters that act as large effective spins excited between quantized states.\cite{Ghosh08} Similar field-induced formation of extended states has been observed as well in the dilute Ising magnet LiHo$_x$Y$_{1-x}$F$_4$, where quantum fluctuations rather than geometric frustration preserve the isolated degrees of freedom.\cite{Ghosh02,Silevitch07,Schmidt14} Here we explore the effects of the Nd doping on these cluster states in GGG.

\begin {figure}
\includegraphics[clip,trim=110 440 120 70,width=250pt]{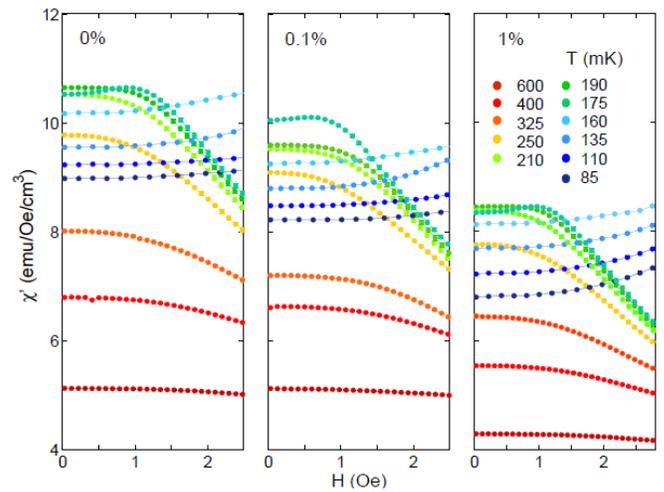}
\caption{Real part of the susceptibility $\chi'$ of GGG:Nd$_x$ in the nonlinear response regime, measured at $f = 5$ kHz as a function of the applied probe field $H$ at a series of temperatures. At elevated $T$, there is clear threshold behavior in the susceptibility and the nonlinear response can be fit to a Brillouin function characteristic of coherent clusters with a large effective spin (see text for details). Below $T = 175$ mK, the cluster dynamics freeze out and the small increase in $\chi$ arises from inductive heating of the susceptometer (and hence the sample) at large $H$.}
\label {fig3}
\end {figure}

We plot in Fig. \ref{fig3} the nonlinear susceptibility at $f = 5$ kHz for the three dopant concentrations over almost a decade in $T$. The response at $f = 15$ Hz is qualitatively similar. There are two distinct regimes in temperature. For $T \geq 175$ mK, the susceptibility is essentially constant at low applied field, followed by a continuous drop once the applied field is above a threshold value. The shape of the susceptibility above the threshold can be described by the derivative of the Brillouin function, $\chi(H)= N_cN_s\mu_BgJ\frac{d}{dH}B_J(y)+\mathrm{const.}$, where $y = \frac{N_s\mu_BgJH}{k_BT}$, $B_J = \frac{2J+1}{2J}\coth(\frac{2J+1}{2J}y)-\frac{1}{2J}\coth(\frac{1}{2J}y)$, and for GGG J $=7/2$ and g = 2. $N_s$ is the number of microscopic spins bound in an individual cluster, $N_c$ is the total number of clusters per unit volume, and the constant term accounts for the susceptibility of spins that are not bound up in clusters. Even in the presence of doping-induced disorder, clusters acting as single large effective spins continue to form. For $T\leq 160$ mK, the spin clusters freeze out and the nonlinear contribution to the susceptibility disappears. The slow rise in $\chi(H)$ at the highest drive fields is most likely due to Ohmic heating that is at worst 25 mK for a 2.75 Oe/5 kHz applied field at lowest $T$ by reference to the linear-response data of Fig. \ref{fig2}.

\begin {figure}
\includegraphics[clip,trim=140 380 120 60,width=250pt]{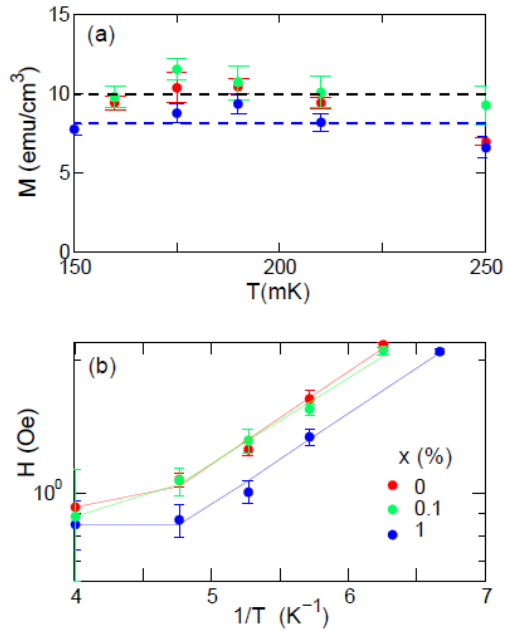}
\caption{Cluster dynamics of GGG:Nd$_x$ at $f = 15$ Hz. (a) Aggregate moment of all spins involved in coherent clusters as a function of temperature $T$ for $x = 0$, 0.1, and 1.0\%. Dashed lines show the mean values for the three doping concentrations. Increased doping reduces the cluster size. (b) Activation field vs. inverse temperature. The Arrhenius form establishes the germane energy scales for cluster spin flips as a function of $x$.}
\label {fig4}
\end {figure}

The fit to the Brillouin form for $T \geq 175$ mK helps specify the nature of the cluster dynamics. We show in Fig. \ref{fig4} the aggregate moment of the spins bound up in coherent clusters at $f = 15$ Hz, as well as the threshold activation field for the onset of the nonlinear response as a function of $T$. The typical cluster size is 150--200 spins, corresponding to 6 to 8 unit cells of the GGG lattice. The moment of the clusters only depends weakly on $T$, until the entire mode freezes out (Fig. \ref{fig3}). There is, however, a definite compositional dependence (Fig.  \ref{fig4}a). The 0 and 0.1\% Nd-doped samples have comparable moments, whereas the 1\% crystal has a net cluster moment reduced by 20\%. The Nd does not act to improve the cluster correlations (by, e.g., acting as nucleation sites for coherent clusters), but instead serves to reduce the disorder in the GGG lattice and more closely demarcates the boundaries of the isolated spin clusters.

While the net moment of the clusters does not vary appreciably with $T$, the threshold field required to excite the coherent response is thermally activated and obeys an Arrhenius form (Fig. \ref{fig4}b) with an energy barrier height of 0.48 $\pm$ 0.05 K. The energy $gJ\mu_BH$ to flip a cluster of 200 spins with $J = 7/2$, $g = 2$ and a threshold field for the nonlinear response $\sim$ 1 Oe is approximately 0.2 K, consonant with the measured barrier height and the scale set by the dipolar coupling. Although the barrier height is equal within error bars for all three Nd concentrations, the threshold field for the 1\% doped sample is reduced by 25\% in comparison to the more lightly doped samples as the smaller clusters more easily change orientation in response to the driving field.

\begin {figure}
\includegraphics[clip,trim=100 260 120 160,width=250pt]{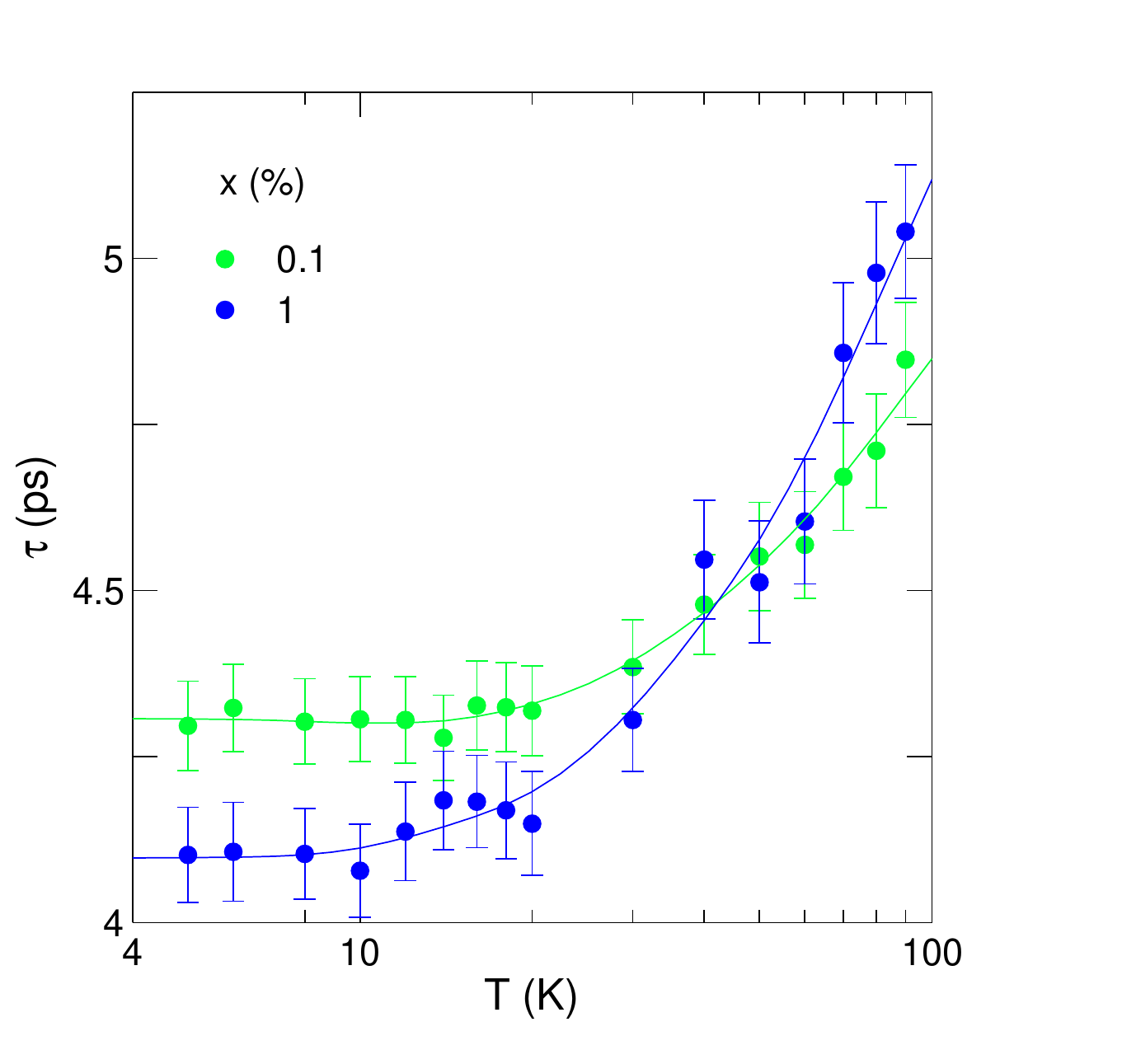}
\caption{Decay time of the $R_1 \to Z_5$ emission line of Nd$^{3+}$ in GGG:Nd$_x$ as a function of temperature for x = 0.1\% and 1.0\%. The faster decoherence time with increased dopant concentration supports the notion of increased frustration.Lines are guides to the eye.}
\label {fig5}
\end {figure}

We show in Fig.~\ref{fig5} the decay time of the optical $R_1 \to Z_5$ emission line of Nd$^{3+}$ in GGG:Nd$_x$ as a function of temperature for $x = 0.1$ and 1\%. Typically, as a system is cooled, the homogeneous linewidth (inversely proportional to the decay time) of a dipole transition decreases with the suppression of the phonon density of states. The contrary trend observed here indicates the availability of a different decoherence mechanism, and while this decreasing trend in decoherence is not seen in other Nd$^{3+}$ garnets like YAG and YAP,\cite{Kushida69,Weber71} it has been reported in Mn$^{4+}$ doped GGG, where it was attributed to spin-spin coupling between the Gd and the dopant ions.\cite{Bulyarskii00} It is important to note that that the excited state lifetime is long in rare earth doped crystals,\cite{Kenyon02} and thus, judging by the extremely fast decay time seen in the data, we can infer that spin-spin interactions are the dominant effect. As the Nd$^{3+}$ ions fill in the Gd vacancies in the GGG lattice, the local environment becomes less disordered, the frustration index increases, and the decoherence time decreases.

Using a Nd doping series to vary the effective degree of disorder in GGG, we demonstrate that increasing the Nd concentration actually reduces the effects of disorder and enhances the signatures of frustration. We posit that the Nd compensates for the disorder arising from the off stoichiometric growth process and, in so doing, permits the intrinsic geometric frustration of the lattice to suppress the formation of an ordered state. The nonlinear susceptibility reveals quantitative aspects of the coherent spin clusters and their tenability with disorder and frustration. In principle, the Nd ions could be marshaled to interrogate locally the isolated cluster dynamics via optical studies at lower $T$.
\end{text}

\begin{acknowledgments}
We thank Y.Feng for assistance with the single crystal x-ray diffraction performed at sector 4 of the APS at Argonne National Laboratory. Work at the University of Chicago was supported by the Department of Energy Office Basic Energy Sciences, Grant No. DE-FG02-99ER45789 and the work at the University of California, Merced was supported by the National Science Foundation, Grant No. DMR-1056860.
\end{acknowledgments}

\end{document}